# Concerning Mössbauer experiments in a rotating system and their physical interpretation

Alexander L. Kholmetskii[1], Tolga Yarman[2], Ozan Yarman[3] and Metin Arik[4]

[1]Department of Physics, Belarus State University, Minsk, Belarus, tel. +375 17 2095482,
e-mail: alkholmetskii@gmail.com
[2]Okan University, Istanbul, Turkey & Savronik, Eskisehir, Turkey
[3]Istanbul University, Istanbul, Turkey
[4]Bogazici University, Istanbul, Turkey

## Abstract

We shortly review different attempts to interpret the results of Mössbauer rotor experiments in a rotating system and particularly we show that the latest work on this subject by J. Iovane and E. Benedetto (Ann. Phys., in press), which claims that the outcomes of these experiments can supposedly be explained via "desynchronization of clocks" in the rotating frame and in the laboratory frame, is inapplicable to all of the Mössbauer rotor experiments performed up to date and thus does not have any significance.

## 1. Introduction

During the past decade, the problem of the correct physical interpretation of the results of Mössbauer experiments in a rotating system gained a significant attention (see, e.g. [1-14]). The increased interest shown to these experiments emerged after our discovery [1] of an extra energy shift (next to the usual energy shift due to the relativistic dilation of time) between a resonant source at the origin of a rotating system and a resonant absorber on the rotor rim, that was originally revealed by our team via the re-analysis of the historical experiment by Kündig [15] based on the stimulating prediction by the second author [16-17], and then confirmed through two separate sets of experiments we subsequently carried out [18-21].

A short review of the recent publications on this subject has been presented in the paper by Iovane and Benedetto [22], where, however, the major part of the contributions by our team (e.g., refs. [5-6, 8-9, 12, 14, 19-21, 31]) is regrettably not mentioned. In this respect, in section 2 we complement the analysis of Mössbauer rotor experiments presented in ref. [22] with our relevant works on the subject, which, in our opinion, allow a better understanding of the existing approaches with regards to the physical interpretation of these experiments.

Further, in section 3, we consider the synchronization procedure between the clock situated at the origin of a rotating system, the clock situated on the rotor rim, and the laboratory clock in order to show that the derived desynchronization effect [22], which allegedly gives an additional contribution to the measured energy shift between emitted and absorbed resonant radiation, is not applicable to any of the experiments on this subject performed to the date, and does not bear any real significance. Thus, in section 4, we emphasize that the problem of the correct interpretation of the results of Mössbauer experiments in a rotating system cannot be reduced to an unaccounted-for desynchronization effects between the clocks in a rotating system and in a laboratory frame – so much so that further search of possible approaches to the explanation of these experiments in the framework of general theory of relativity (GTR) is required.

## 2. Mössbauer experiments in a rotating systems and the extra energy shift between emission and absorption lines: experimental and theoretical approaches

The original series of Mössbauer rotor experiments (e.g., [15, 23-27]) had been carried out in the 1960s soon after the discovery of the Mössbauer effect with an aim to verify the relativistic ef-fect of time dilation under laboratory conditions, which manifests itself via the second order Doppler shift between the resonant lines of a source at the origin of a rotating system and absorber on the rotor rim. Among these experiments, special attention should be accorded to the experiment by Kündig [15], who, for the first time, realized a linear Doppler modulation between the resonant lines of the source and the absorber on a spinning rotor, so that the measured position of the resonant line on the energy scale was not influenced by random vibrations, which are always present in a rotating system. Thus, from the methodological viewpoint, Kündig′s experiment [15] should be considered as the most reliable among other experiments of the 20$^{th}$ century on this subject [23-27]. Here, we remind that Kündig reported the confirmation of the standard expression for the second order Doppler shift, i.e.,

$$z = -\frac{1}{2}\frac{\omega^2 r^2}{c^2},$$ (1)

where $z$ stands for the relative energy shift between emission and absorption lines, $\omega$ is the angular velocity of the resonant absorber, $r$ is its radial coordinate, and $c$ in the light velocity in vacuum.

Under these conditions, the disclosure of computational errors by Kündig in the data processing [15], as revealed in our paper [1], indeed deserves considerable attention.

We must, in relation, note that the authors of [22] incorrectly presented the result of our reanalysis as[1]

$$\frac{k}{2} = 0.68 \pm 0.03.$$ (2)

In this respect, we emphasize that the actual result derived in ref. [1] is given by the expression

$$k=1.192\pm0.011,$$ (3)

which in our subsequent publications, had been modified to the inequality

$$\frac{k}{2} > 0.60$$ (4)

due to the presence of possible systematic errors affecting the measurement results and reducing the actual value of $k$, and which could not be exactly evaluated in the experiment by Kündig.

Concerning eq. (2), which is mentioned in [22] as the result of our re-analysis of the experiment by Kündig; it represents, as a matter of fact, the outcome of our own Mössbauer experiments in a rotating system [18], carried out in Minsk in 2008, and later corrected to the value [19]

$$\frac{k}{2} = 0.66 \pm 0.03,$$ (5)

due to updated information [28] about the Debye temperature of resonant absorbers used in our measurements.

What is more, the authors of ref. [22] even did not refer to the most precise experiment by out team conducted in Istanbul in 2014 [20, 21], which revealed

$$\frac{k}{2} = 0.69 \pm 0.02.$$ (6)

Having missed the essential information presented above, the authors of [22] further claim that the experiment by Kündig "… *has been recently reanalyzed and replied* [8-10]…" (which now are respectively refs. [3], [29] and [1]). This again is not correct, because the mentioned papers by Friedman et al. [3], [29] had been published already after our re-analysis [1] of the experiment by Kündig, and after the performance of our experiment [18] mentioned above, yielding the result (2). As for the experiments [3], [29], they were aimed to validate the claim

---

[1] For the convenience of the readers, we reproduce the definition of the coefficient $k$ given in ref. [22] in the form

$$z = -k\frac{1}{2}\frac{\omega^2 r^2}{c^2},$$

which differs by a factor of 2 from the definition of this coefficient, adopted in our papers on the subject.

that eq. (3) can be understood within the hypothesis about the existence of a maximal acceleration in nature, as is supposed to be the case in the generalized relativistic kinematics [2, 3] by Friedman et al.

We remind that the hypothesis about a limited acceleration in nature had been advanced in the 20$^{th}$ century – see, e.g., ref. [30], where its assumed value is

$$a_m = c^2/l_p \approx 5.5 \times 10^{51} \, m/s^2, \tag{7}$$

with $l_p$=1.62×10$^{-35}$ m being the Planck length.

Contrary to the fundamental estimation (7), Friedman et al. assumed *ad hoc* that the actual value of a posited maximal acceleration is much smaller than (7), and its value, they deemed, can be determined in the Mössbauer rotor experiment through the measurement of the coefficient $k$, which should depend on $a_m$ via the relationship [2, 3]

$$k = 1 + \frac{2c^2}{ra_m}. \tag{8}$$

Hence, using our re-estimation of the result by Kündig (3), and taking the radial coordinate of the resonant absorber $r$=9.3 cm in the Kündig experiment, Friedman et al. obtained

$$a_m \approx 1.01 \times 10^{19} \, \text{m/s}^2. \tag{9}$$

We see that the estimation (9) is many orders of magnitude smaller than the purported fundamental limit of maximal acceleration (7). At the same time, one can agree with Friedman et al. that only experiments can shed light on the actual value of $a_m$, if it exists.

Thus, in order to verify the result (9), recently Friedman et al. carried out their own Mössbauer experiments in a rotating system [29], where they applied resonant synchrotron radiation of the European Synchrotron Radiation Facility (ERSF). Their measurements yielded

$$a_m \approx 1.2 \times 10^{17} \, \text{m/s}^2, \tag{10}$$

which, though, is almost two orders of magnitude smaller than their original estimation (9) made via eqs. (8) and (3).

A drastic disagreement between two estimations (9) and (10) with respect to the value of $a_m$ nevertheless does not, in the opinion by Friedman et al., invalidate their approach, but rather indicates that the maximal acceleration, if it exists, cannot be considered as a universal quantity.

Here, it is worth mentioning that in our analysis of the experiment by Friedman et al. [29] carried out in ref. [31], we indicated the presence of unaccounted systematic errors in the experiment [29], so that its result is anyway inconclusive.

Thus, considerable methodological improvements are required for the further development of the experimental approach by Friedman et al. [29] in order to obtain actual information about the limited value of $a_m$, provided that the latter exists at all.

We add that the authors of [22] classified the experiment by Friedman et al. [29] as "a very interesting" undertaking, but seem to have forgotten to refer our paper [31], where we already presented a crucial analysis of said experiment.

A similar omission of our contribution to the subject takes place in the review of [22] with respect to one more paper that aims to explain the result of the experiment by Friedman et al. [29] on the basis of a so-called "time-dependent Doppler shift" [13]. The authors of [22] posited such a "time-dependent Doppler shift" as a prevalent idea capable of explaining the results of Mössbauer experiments in a rotating system, but they completely hid from view our contribution [14], where we indicated at least two principal errors in ref. [13], which totally invalidate the ideas of that paper.

Finally, completing the review of Mössbauer rotor experiments and different attempts of their interpretation, the authors of [22] refer to the approach by C. Corda [7, 11] to explain the experimental results (5) and (6) via the effect of synchronization of a clock at the origin of a rotating system and a laboratory clock. According to Corda, this so-called synchronization effect gives an additional contribution to the measured energy shift and yields the total shift

$$z = -\frac{2}{3} \frac{\omega^2 r^2}{c^2}, \tag{11}$$

seemingly in a perfect agreement with the measurement results (5), (6). The authors of [22] classify this approach by Corda as the "most successful" among other ones, but unfortunately again miss our contributions [8, 12] where we explained in sufficient details that the purported effect of the synchronization of clocks − even should it ever exist − could not in any way be measured via Mössbauer rotor experiments, and thus ought to be ignored altogether. Furthermore, the authors of [22] reproduce the assumption by Corda [11] that eq. (11) "...*represents the total energy shift that is detected by the resonant absorber as it is measured by an observer located in the detector of γ-quanta. Instead,... the shift detected by an observer located in the resonant absorber, is given by standard Eq. (1).*" (which is also eq. (1) in the present paper). However, in ref. [12], we have shown that these repetitious claims are devoid of sense, and they lead to a contradiction with classical causality where two observers, as defined above, should fix different total numbers of detected γ-quanta after the completion of any measurement run. Nevertheless, the authors of [22] again chose to overlook our papers [8, 12] which disclose the errors by Corda in [7, 11].

Even so, starting with the self-titled "interesting Corda approach" Iovane and Benedetti formulate the goal of their paper [22] to explore, how "...*the term he has found is linked to the difference between coordinate and local velocity of light*".

In the next section, we analyze the results obtained in [22] and show that:

1. The problem which Iovane and Benedetti embark to solve has no relationship with the actual configuration of all known Mössbauer rotor experiments and thus it has no practical significance in further analysis of these experiments.

2. The only useful implication of the results presented in [22] is the explicit disproof of the approach by Corda, in spite of the claim by Iovane and Benedetti that they are only aimed to its further development.

### 3. Mössbauer experiments in a rotating systems and "desynchronization" of clocks

In the analysis of Mössbauer rotor experiments, the authors of [22], following Corda [7], use the Langevin metric

$$ds^2 = \left(1 - \frac{r'^2\omega^2}{c^2}\right)c^2dt'^2 - 2\omega r'^2\,d\theta'dt' - dr'^2 - r'^2\,d\theta'^2 - dz'^2 , \qquad (12)$$

in the cylindrical coordinates. We remind that eq. (12) directly stems from the transformation between an inertial frame and a rotating frame, adopted in the form [32]

$$t = t',\ r = r',\ \theta = \theta' + \omega t',\ z = z', \qquad (13\text{a-d})$$

where the non-primed quantities refer to the inertial (laboratory) frame, while the primed quantities are referred to the rotating frame.

Using the metric expression (12), Iovane and Benedetti considered the propagation of photons in the radial direction of the rotating system, and further defined the proper time of the absorber next to the proper time of the detector, and they finally derived the expected energy shift between emission and absorption lines in Mössbauer rotor experiments involving the so-called effect of the "desynchronization of clocks" that they have introduced.

Below, we skip a detailed analysis of their approach so as to emphasize the most important fact, which makes the results obtained in [22] totally inapplicable to any real experimental setups realized in Mössbauer rotor experiments. This conclusion stems from the fact that Iovane and Benedetti considered only the case where the photons propagate in the radial direction of the rotating system, where $\theta'$=*constant*, and

$$d\theta' = 0 . \qquad (14)$$

However, the fact remains that in all known Mössbauer experiments in a rotating system implemented up to date (including [15, 18-21, 23-27]), the resonant γ-quanta propagate over the rotor along the straight line, joining a spinning source and a detector as seen by a laboratory observer (Fig. 1). In such a case, instead of the equality $\theta'$=*constant*, we have $\theta$=*constant*. Thus, putting $d\theta$=0 in eq. (13d), we obtain

$$d\theta' = -\omega dt' \qquad (15)$$

instead of eq. (14) used in [22] in the analysis of implications of eq. (12).

At the same time, it is obvious that the replacement of eq. (14) by eq. (15) drastically changes the result of the analysis. For example, instead of the equation

$$ds^2 = \left(1 - \frac{r^2\omega^2}{c^2}\right)c^2dt^2 - dr^2$$

obtained by Iovane and Benedetti with eq. (14) and $dz'=0$ for the radial propagation of light in the rotating frame (see eq. (12) of [22]), we derive with eq. (15) and $dz'=0$ (i.e., for the radial propagation of light in the laboratory frame)

$$ds^2 = c^2dt'^2 - dr'^2. \qquad (16)$$

Hence, at $ds=0$ for light, we arrive at the trivial result

$$\frac{dr'}{dt'} = \frac{dr}{dt} = c, \qquad (17)$$

or

$$t = r/c. \qquad (18)$$

Comparing this equation with the corresponding equation (15) of [22], i.e.,

$$t = \frac{r}{c} + \frac{\omega^2 r^3}{6c^3} + O\left(\frac{\omega r}{c}\right)^5, \qquad (19)$$

obtained with eq. (14) in the case of the assumed propagation of photons in the radial direction of the rotating frame, we can, with some stretch of the imagination, understand its physical meaning. Effectively, at least in principle, one can force photons to propagate in the radial direction of the rotating frame, where the equalities $d\theta'=dz'=0$ used by Iovane and Benedetti in [22] could be actually fulfilled. For this purpose, one has to connect the source and the absorber by a very thin guide for resonant gamma-quanta (when such guides might be invented in the future), which should be rigidly attached to the rotor. In such a case, for a laboratory observer, the photons inside the rotating guide would hence be coerced to propagate along a curved path and would reach the radial coordinate $r$ of the resonant absorber later than the photons which propagate along a straight line of the laboratory frame, insofar as making understandable the difference between eqs. (19) and (18). Yet, in the absence of any such guide of resonant $\gamma$-quanta, the whole premise of a photon following a "curved path" above the rotor in virtually empty space falls apart.

Therefore, at the present time, equation (19) remains unreal, for it has no relationship to any of the known Mössbauer rotor experiments performed up to date, and should be denied. Thus the subsequent analysis in sections 3-5 of [22], which is essentially based on eq. (19), should be denied, too.

There is only one equation (28) in section 5 of [22], which is applicable to all real Mössbauer rotor experiments, where Iovane and Benedetti show that the proper time for the detector located outside the rotor coincides with the coordinate time. However, this result is strongly at odds with the opposite claim by Corda in refs. [7, 11]. Thus, as a matter of fact, the equation (28) of [22] serves to disprove the entire approach by Corda, even if Iovane and Benedetti do not pay any attention to this factuality. The fact remains that, in aspiring to save Corda's approach, they have unwillingly come to destroy it.

Concurrently, we notice that the coincidence of the proper time for the detector with the coordinate time makes eq. (18) to be trivial and indicates that, in the case of propagation of light along a straight line of the laboratory frame, there is no desynchronization effect derived in ref. [22] with the irrelevant eq. (19).

## 3. Conclusion

In this paper, we complemented the review presented in ref. [22] with respect to Mössbauer rotor experiments, where several papers of our team were omitted by Iovane and Benedetti.

We further pointed out that the results derived in [22] cannot be applied to the explanation of any known experiments on this subject, insofar as Iovane and Benedetti considered only the unreal case of the propagation of photons in the radial direction of a rotating frame, whereas the actual propagation of $\gamma$-quanta happens along the straight lines of the laboratory frame as shown in Fig. 1.

Therefore, the results of ref. [22] could be topical only in a futuristic scenario, when guides for resonant $\gamma$-quanta might be invented and applied to Mössbauer experiments in a rotating system.

Thus, at this stage, the problem of correct interpretation of these experiments within GTR still remains open.

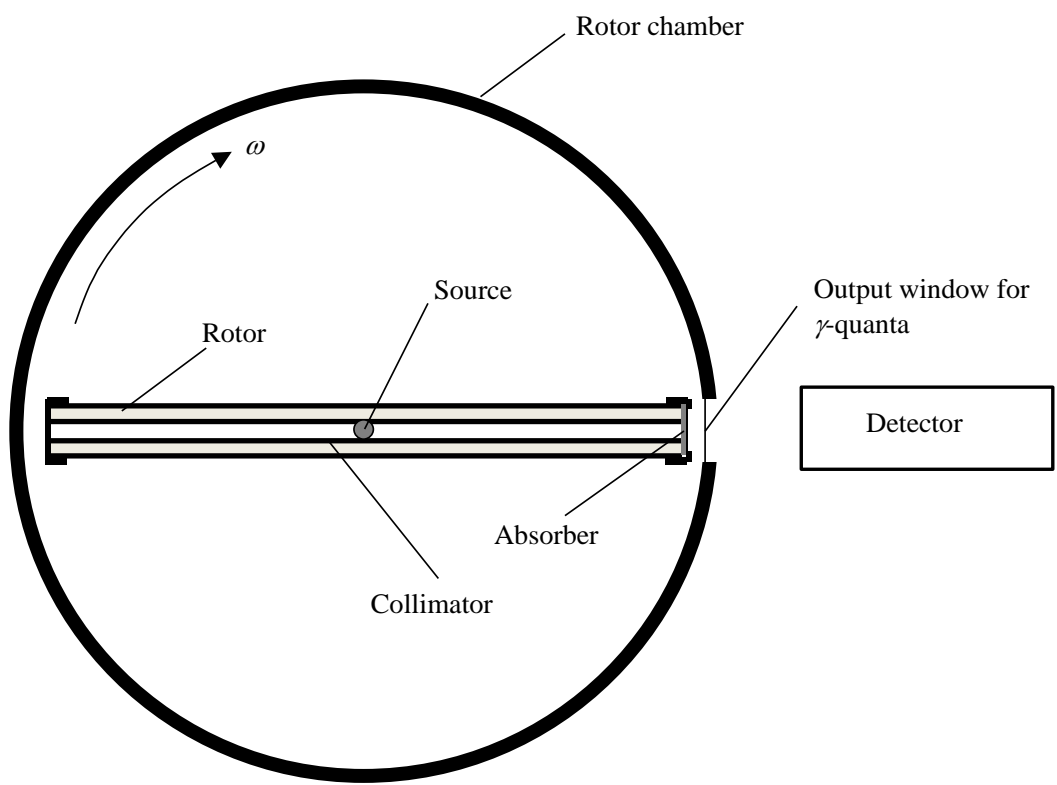

Figure 1. General scheme of the Mössbauer rotor experiment. A source of resonant radiation is located on the rotor axis; an absorber is fixed on the rotor rim, while a detector of $\gamma$-quanta is placed outside the rotor system. $\gamma$-quanta emitted by the source and passing across the absorber are detected at the time moments, when source, absorber and detector are aligned in a straight line, as is seen by a laboratory observer.